\documentclass[12pt,preprint]{aastex}

\newcommand{\Rs}{R_{\odot}}

\shorttitle{Radius variation and solar subsurface stratification}
\shortauthors{Lefebvre et al.}

\begin{document}

\title{Impact of a radius and composition variation on the solar subsurface layers stratification}

\author{S. Lefebvre, P. A. P. Nghiem and S. Turck-Chi\`eze}
\affil{Irfu/SAp, CEA/DSM and Laboratoire AIM: CEA- CNRS - Universit\'e Paris Diderot,\\ 91191 Gif sur Yvette, France}
\email{sandrine.lefebvre@cea.fr}
           
\begin{abstract}

Several works have reported changes of the Sun's subsurface stratification inferred from $f$-mode or $p$-mode observations. Recently a non-homologous variation of the subsurface layers  with depth and time has been deduced from $f$-modes. Progress on this important transition zone between the solar interior and the external part supposes a good understanding of the interplay between the different processes which contribute to this variation. This paper is the first of a series where we aim to study these layers from the theoretical point of view. For this first paper, we use solar models obtained with the CESAM code, in its classical form, and analyze the properties of the computed theoretical $f$-modes. We examine how a pure variation in the calibrated radius influences the subsurface structure and we show also the impact of an additional change of composition on the same layers. Then we use an inversion procedure to quantify the corresponding $f$-mode variation and their capacity to infer the radius variation. We deduce an estimate of the amplitude of the 11-year cyclic photospheric radius variation.

\end{abstract}

%-------------------------------------------------

\section{Introduction}
     \label{S-Introduction} 
     
Helioseismology has been extremely useful to probe the internal structure of the Sun but two crucial regions need to be improved for a good understanding of the time evolution of the solar activity: (1) the solar core for a proper description of the transport of momentum implied by  rotation,  gravity waves and magnetic field along the evolution and (2) the subsurface layers which correspond to the transition region between large  and small scale dynamics evolution.

\begin{figure*}[htbp]     
  	%\centerline{\includegraphics[width=0.5\textwidth,clip=]{fig1a.eps}}
   \centerline{\includegraphics[angle=0,width=16cm]{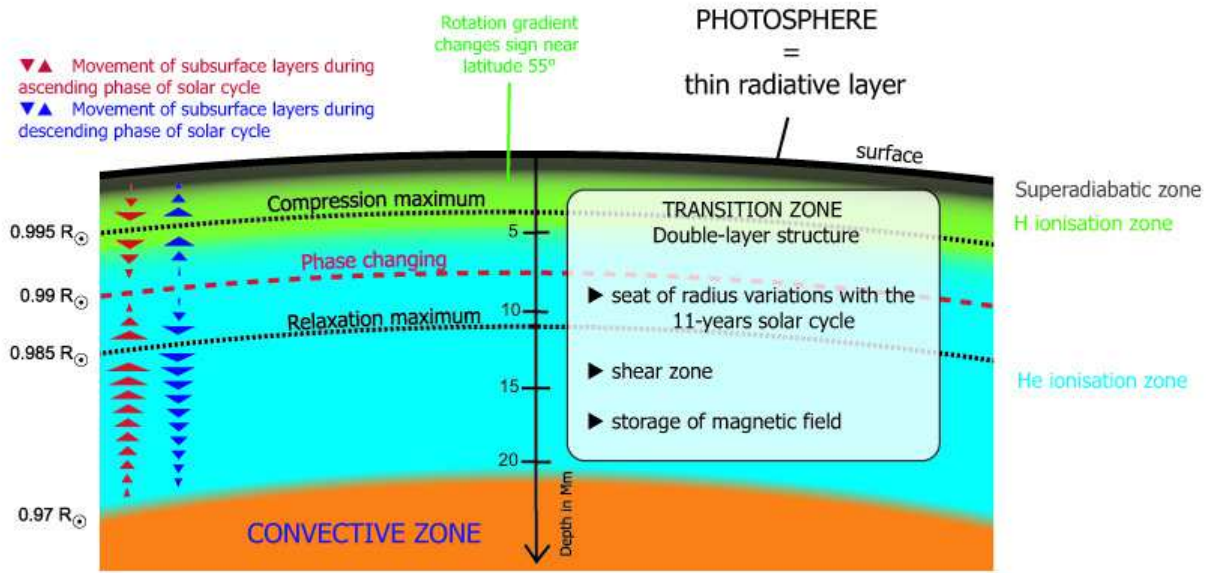}}
   \caption{Schematic view of the leptocline (non-scale scheme).}
   \label{F-lepto}
\end{figure*}

This transition zone is important to study because it couples the internal dynamics to the dynamics of  the solar atmosphere and it plays  a crucial role in the emergence of the space weather science (see for example \citet{Rozelot06}). Figure \ref{F-lepto} sketches a schematic view of these subsurface layers above 0.96 $\Rs$, which are called the \textit{leptocline} region (from the greek ``leptos''= thin, ``klino"= tilt). This term was proposed by \citet{Godier01}, who computed the solar oblateness and showed a curvature change in this zone which was interpreted as the presence of a double layer. Just below the surface, there is a change in radial and latitudinal rotation \citep{Basu99,Corbard02} and the treatment of the superadiabiatic region supposes a proper description of the convection (presently described by the mixing length parameter)  and of detailed molecular and atomic opacity calculations. One needs also to add the turbulent pressure and the emergence of local magnetic field to these processes, then below, hydrogen and helium pass from neutral to partially ionized and then totally ionized in a region where the magnetic pressure cannot be ignored. The mean and varying magnetic field amplitudes have been tentatively extracted by \citet{Nghiem06} from the analysis of the low degree acoustic modes.%, it is also the seat where the radius variation with the 11-year cycle can be observed. 
The proper interaction between these different physical processes is not yet included in stellar evolution models and could lead to some variation of the solar radius along the solar cycle.

The observation of these subsurface layers is not so easy. The difficulty comes from the fact that the high-degree $p$-modes \citep{Korzennik04} have a small lifetime and are largely perturbed by the turbulent motions and the emerging magnetic field. But important progress has appeared recently  thanks to the analyses of long series of seismic data. \citet{Rabello08}  estimate the variations of the high-degree $p$-modes frequencies over the solar cycle and  using ring diagram analysis  \citet{Basu07}  extract now latitudinal and temporal variations of  the sound speed or the density along the Hale solar cycle (at least its last half cycle). 

\clearpage
% Figure 1
\clearpage

In parallel, \citet{Lefebvre05} and \citet{Lefebvre07} reported changes with the solar cycle of the solar subsurface stratification, inferred from inversion of SOHO/MDI $f$-mode frequencies. They notice a different behavior for the layers around 0.99 $\Rs$. Between 0.97 and 0.99 $\Rs$, it seems that the position of the layers varies in phase with the solar cycle, whereas it appears opposite in the upper part, above 0.99 $\Rs$, where the variability is in antiphase. 
So,  it seems that the most external layers of the Sun shrink during the ascending phase of the solar cycle and relax after the maximum, but these changes are not uniform with depth. At the surface, this result  is coherent with the observation of the photospheric radius variations. The observed series of ground-based measurements  \citep{Laclare96,Reis03} and the measurements aboard stratospheric balloons \citep{Sofia94} suggest a reduction of the solar photospheric radius at the maximum of the cycle. The balloon variation estimates are much smaller than the ground-based observations, polluted by the atmospheric turbulence. They are consistent with results from \citet{Kuhn04} who reported no evidence of solar-cycle visible radius variations between 1996 and 2004 larger than 7 mas. But they disagree with computations of \citet{Sofia05}. These authors predict also non-homologous subsurface stratification changes with the 11 year solar cycle but with an amplification up to a factor 1000 from the depth at 5 Mm to the surface. It leads to a variation of the solar radius up to 600 km along the cycle  as shown in Fig. 2 of \citet{Lefebvre07}. Such a variation seems extremely large.

So the study of the detected  $f$-modes is very useful  in the present context even the interpretation of $f$-modes has appeared puzzling \citep{Dziem01,Antia03,Dziem04} due to some apparent lack of sensitivity of these modes at the real surface. We develop here a theoretical approach  to validate  the $f$-modes inversion procedure used previously and applied here to some specific cases that we know perfectly.
This paper is the first of a series where we estimate the impact of a change of radius and composition on the subsurface layers and on the $f$-mode frequencies using classical solar models including a detailed microscopic description of these layers. In the present work, we do not try to justify the origin of the variation of the radius. The next step will be to develop more complex models including magnetic field and differential rotation. Our general aim  is to properly qualify the variabilities of $f$-modes and solar radius that we will study with the SDO and PICARD missions (Kosovichev et al. 2007 and Thuillier, Dewitte \& Schmutz 2006, respectively).%\citep{Kosovichev07}  \citep{Thuillier06}. 
We will also investigate the capability of the $f$-modes to estimate the solar photospheric radius variability. This study will contribute also to clarify the notion of ``solar radius" to bridge two different communities and obtain without ambiguity a unique definition of this fundamental quantity \citep{Haberreiter08}.

After a description of our study (Section \ref{S-Context}), we describe in Section \ref{S-model}, the models used and examine the subsurface layer changes on different variables produced by a change in the solar radius and composition and we calculate the corresponding $f$-mode frequencies. Section \ref{S-inversion} is devoted to the use of these frequencies to infer the change in the position of the subsurface layers. The validation of the procedure, an estimate of the present solar cyclic radius variation and the perspectives of this work will conclude the paper in Section \ref{S-discussions}.

\section{A new estimate of the solar cycle radius change using models and f-modes predictions}
	\label{S-Context}
	
The present theoretical work focus on the solar layers located above 0.96 R$_\odot$. The previous approach of \citet{Li03} tried to interpret the behavior of the seismic acoustic observations by introducing some magnetic effect in solar models to predict the radius variation over the cycle. In the present work, we concentrate on the $f$-mode frequencies. We use known models and perform inversions that we can verify and  compare with  the present observational $f$-mode frequency variations. For this first paper, we stay in the classical approach and only focus on basic quantities which may vary along the solar cycle without introducing the physical process  at the origin of the corresponding variations. 

The next objectives will be to introduce the different magnetohydrodynamical actors \citep{Mathis05}.  
Turbulence, rotation and the local effect of magnetic field with a decent topology \citep{Duez07} must be introduced together in a more sophisticated way than it is generally done. But we would like first to study the potential of the $f$-modes in some well known cases and then show that it is possible to deduce an order of magnitude of the solar radius variation from the present variation of the observed $f$-mode frequencies.

The general idea is to progress step by step on the sensitivity of the $f$-modes to the complex physics of the subsurface layers using the same formalism than the one used in \citet{Lefebvre05}. In the present paper we do not yet impose any kind of magnetic field like in \citet{Li03} or \citet{Mullan07} but we estimate the local changes to decouple the origin of the different effects. 
We examine  the theoretical $f$-mode frequency changes coming from known solar models which  mimic a simple expansion or a change in composition. We deduce, for the first time from  $f$-modes, an estimate on the quality of extraction of the photospheric radius change and an order of magnitude of its Hale cycle change.

\section{Model computations}
			\label{S-model}

We use the CESAM evolution code \citep{Morel97} to calculate several models of chosen solar radius $R$, luminosity $L$ and surface abundance $\left(Z/X\right)$. The different computed cases are organized in 3 sets, and listed in Table \ref{T-table_models}. 
Our reference model $M_1$ \citep{Turck01,Couvidat03} with $\Rs=6.9599\times10^5 \rm km$, $L_{\odot}=3.8460\times10^{33} \rm erg/s$ and $Z/X_{\odot}=0.02447$ is the seismic solar model (SeSM) built to reproduce the observed sound speed profile. 

Set 1 is composed of models $M_2$ and $M_3$  converged with a radius fixed respectively at $\frac{R_2}{\Rs}=1.0002$ and $\frac{R_3}{\Rs}=0.9998$ and the composition fixed at the value of the reference model, but where the mixing-length parameter $\alpha$ is a free parameter (the luminosity is not constrained in models $M_2$ and $M_3$). The radius difference of $2\times10^{-4}$ corresponds to a difference of about 140 km. 

Set 2 is composed of six models $M_i$, for $i=4...9$ built with the two free parameters $Y_i$ the initial composition in helium and $\alpha$; in these models we impose a change in the radius and luminosity, and the same ratio $Z/X_{\rm surf}$ of the final superficial chemical composition than that of model $M_1$. 

Finally set 3 consists of five models $M_i$, for $i=10...14$ which differ from set 2 by the value of $Z/X_{\rm surf}$, and includes a model where $R=\Rs$ and $L=L_{\odot}$ ($M_{14}$). A change of composition is interesting to study because the presence of varying magnetic field in the subsurface layers may lead to an apparent change of hydrogen/helium thermodynamic conditions in solar models with some impact on the density and pressure of the subsurface layers.

The goal was to mimic variations of the radius and luminosity which could be extrapolated to variations during the 11-year solar activity cycle. For instance, $M_4$ and $M_9$ have a luminosity varying in the same way than the radius, whereas $M_6$ and $M_7$ present luminosity and radius varying in an opposite way. The relative variations $2\times10^{-4}$ for the radius are chosen larger than the supposed observed ones in order to avoid numerical accuracy problems and to better see their influence on the subsurface layers. In the different cases, the models converge to get the calibrated radius within a precision of 10$^{-5}$ that means within 7 km. The relative change in the luminosity of $10^{-3}$ is similar to the observed solar irradiance variation during the 11-year cycle. The way we introduce it in solar models leads generally to a variation of luminosity on the production of energy instead on the superficial layers. So we see only an indirect effect of the luminosity change on the subsurface layers through changes in temperature and radius, their effect is not negligible for models 10 and 13.

\subsection{Changes of the subsurface layers}
We estimate how the subsurface layers react to the induced changes. For that, 
we estimate  the differences of the following variables between the different models and the seismic model\footnote{Note that each fractional radius $x_i$ is reported to the reference radius, i.e. $x_i^{new}=x_i~R_i/R_{ref}$ to get all models refered to the reference model.}: the mass $m$, the temperature $T$, the density $\rho$, the pressure $p$, the sound speed $c$, the adiabatic exponent $\Gamma_1=(\frac{d\ln p}{d\ln \rho})_{\rm adia}$, the density scale height $H_{\rho}=-\left(\frac{d\ln \rho}{dr}\right)^{-1}$, the pressure scale height $H_p=-\left(\frac{d\ln p}{dr}\right)^{-1}$, the radiative temperature gradient $\nabla_{\rm rad}=\frac{d\ln T}{d\ln p}$, the real temperature gradient $\nabla_{\rm real}={\rm min}(\nabla_{\rm rad},\nabla_{\rm adia})$, the Rosseland opacity coefficient $\kappa$ and the gravitational energy $Eg$.
%\begin{itemize}
%	\item the mass $m$ 
%	\item the temperature $T$
%	\item the density $\rho$
%	\item the pressure $p$
%	\item the sound speed $c$ 
%	\item the adiabatic exponent $\Gamma_1=(\frac{d\ln p}{d\ln \rho})_{\rm adia}$
%	\item the density scale height $H_{\rho}=-\left(\frac{d\ln \rho}{dr}\right)^{-1}$
%	\item the pressure scale height $H_p=-\left(\frac{d\ln p}{dr}\right)^{-1}$
%	\item the radiative temperature gradient $\nabla_{\rm rad}=\frac{d\ln T}{d\ln p}$ 
%	\item the real temperature gradient $\nabla_{\rm real}={\rm min}(\nabla_{\rm rad},\nabla_{\rm adia})$
%	\item the Rosseland opacity coefficient $\kappa$
%	\item and the gravitational energy $Eg$.
%\end{itemize}

\clearpage
\begin{table}[htbp]
\begin{center}
\caption{List of the calculated models.} 
\label{T-table_models}
\begin{tabular}{cccccccc}
\tableline\tableline
Model & $X_i$ & $Y_i$ & $\alpha$ & $R/\Rs$ & $L/L_{\odot}$ & $\left(Z/X\right)_{\rm surf}$ & $T_{eff} [K]$ \\
\tableline
\multicolumn{8}{c}{\textbf{Seismic model}}\\
1 &0.70642 &0.27468 &2.03990 &1.0000 &1.0000 &0.02447 &5777.52 \\
\multicolumn{8}{c}{\textbf{Set 1}}\\ % Modeles avec uniquement dR
2 &0.70642 &0.27468 &2.03751 &1.0002 &1.0000 &0.02447	&5776.97 \\
3 &0.70642 &0.27468 &2.04260 &0.9998 &1.0001 &0.02447 &5778.30 \\
\multicolumn{8}{c}{\textbf{Set 2}}\\ % Modeles avec dR,dL et Z/X =
4 &0.70632 &0.27478 &2.04033 &1.0002 &1.0010 &0.02447	&5778.40 \\
5 &0.70642 &0.27468 &2.03757 &1.0002 &1.0000 &0.02447	&5776.98 \\
6 &0.70651 &0.27458 &2.03463 &1.0002 &0.9990 &0.02447	&5775.52 \\
7 &0.70634 &0.27476 &2.04508 &0.9998 &1.0010 &0.02447	&5779.56 \\
8 &0.70644 &0.27466 &2.04214 &0.9998 &1.0000 &0.02447	&5778.10 \\
9 &0.70653 &0.27457 &2.03937 &0.9998 &0.9990 &0.02447	&5776.67 \\
\multicolumn{8}{c}{\textbf{Set 3}}\\ % Modeles avec dR,dL et Z/X different
10&0.69772 &0.28186 &2.06148 &1.0002 &1.0010 &0.02676	&5778.42 \\
11&0.69791 &0.28166 &2.05575 &1.0002 &0.9990 &0.02676	&5775.53 \\
12&0.69774 &0.28184 &2.06624 &0.9998 &1.0010 &0.02676	&5779.57 \\
13&0.69793 &0.28164 &2.06048 &0.9998 &0.9990 &0.02676	&5776.69 \\
14&0.69783 &0.28175 &2.06098 &1.0000 &1.0000 &0.02676 &5777.00 \\
\tableline
\end{tabular}
\tablecomments{Model 1 is the reference model (Turck-Chi\`eze {\it et al.} 2001; Couvidat, Turck-Chi\`eze and Kosovichev, 2003), $X_i$ and $Y_i$ are respectively the initial composition in hydrogen and helium, $\alpha$ is the mixing-length parameter, $R$ and $L$ are the radius and luminosity, $Z/X_{\rm surf}$ is the final superficial abundance and $T_{eff}$ the effective temperature.}
\end{center}
\end{table}
\clearpage

\clearpage	
% Figure 2
\begin{figure*}[htbp]
	\centerline{\includegraphics[angle=-90,width=16cm]{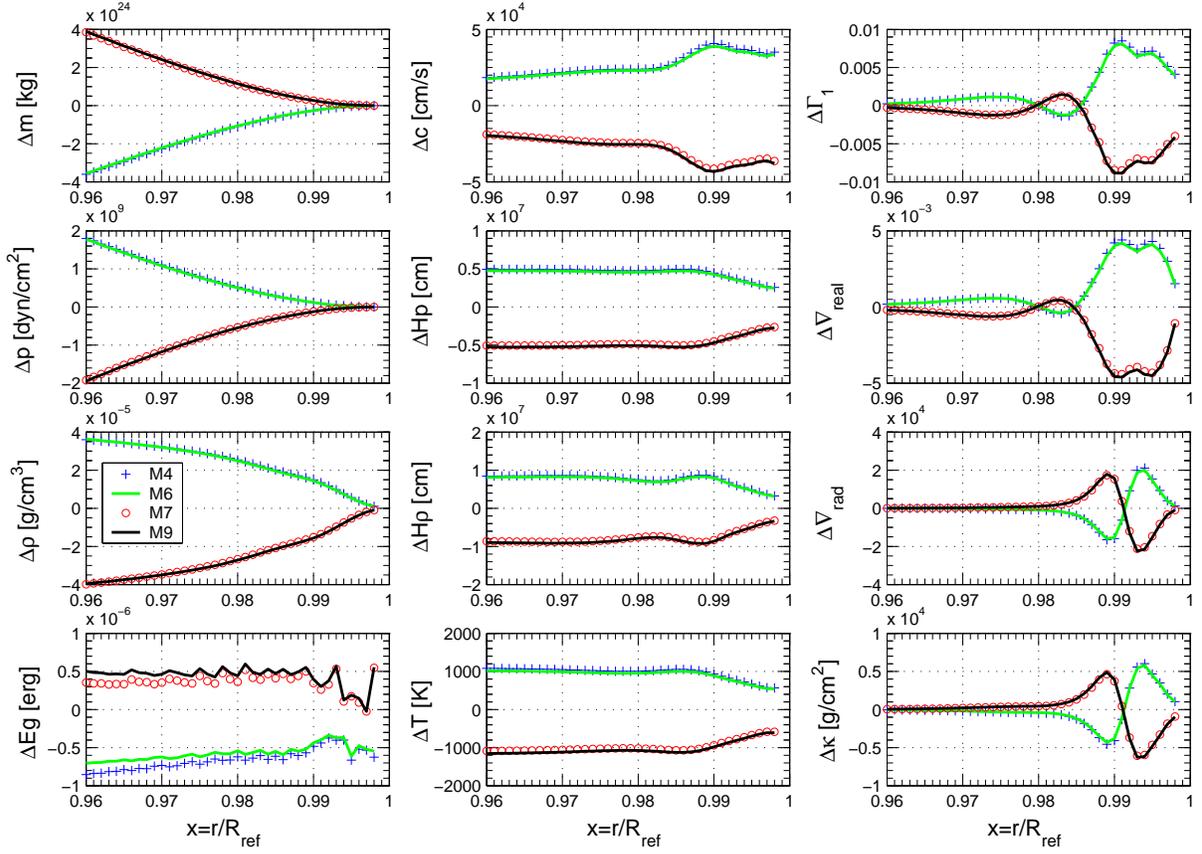}}
	\caption{Difference for the physical variables between models of set 2 calibrated with different combinations of radius and luminosity (see text) and the seismic solar model. The curves of set 1 are exactly superimposed with those of set 2. $M_2$ (or $M_3$) and $M_5$ (or $M_8$) looks like {$M_4$ and $M_6$} (or {$M_7$ and $M_9$}).}
	\label{F-param}
\end{figure*}

% Figure 3
\begin{figure*}[htbp]
	\centerline{\includegraphics[angle=-90,width=16cm]{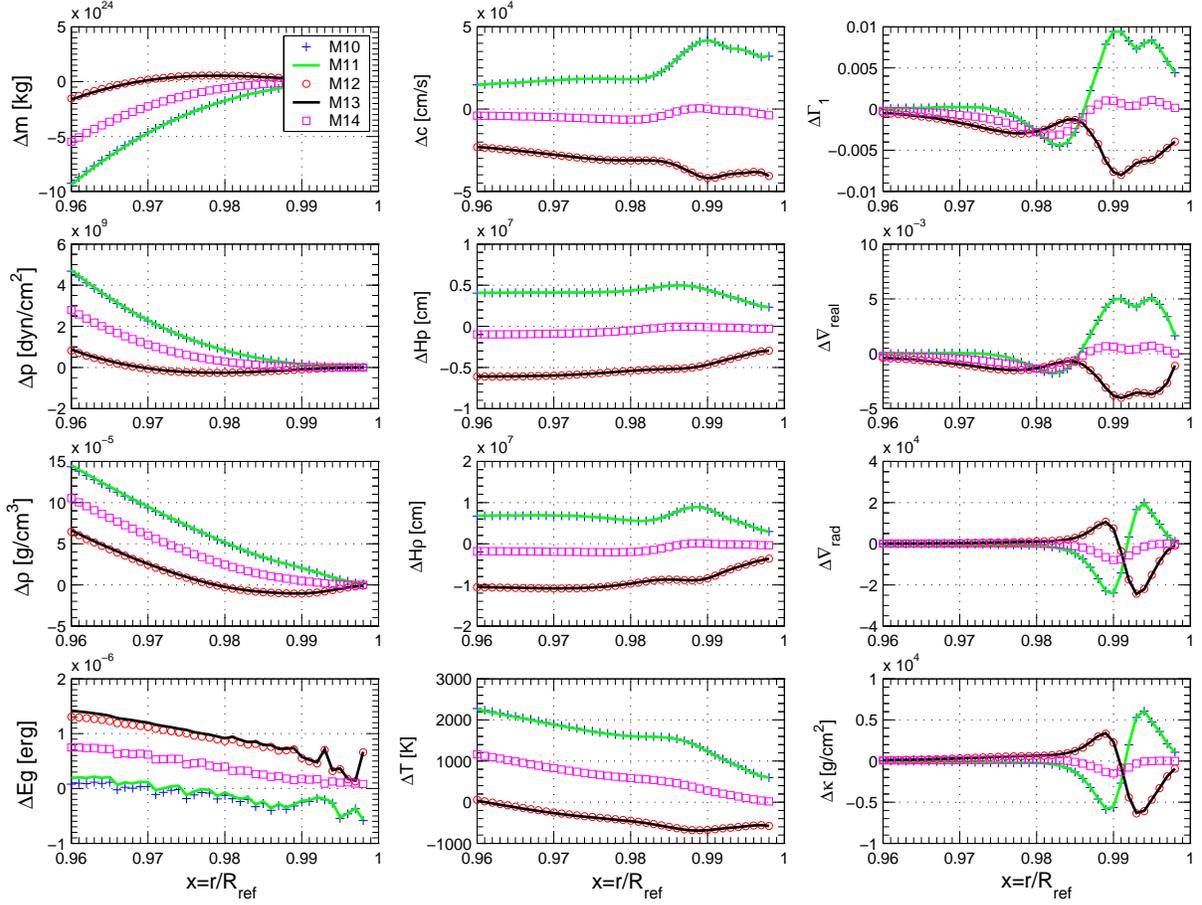}}
	\caption{Difference for the physical variables between models of set 3 and  the seismic solar model (same principle than in Figure \ref{F-param} but including an additional effect in the composition).}
	\label{F-param2}
\end{figure*}
\clearpage

Figures \ref{F-param} and \ref{F-param2} illustrate the changes with depth of these variables in comparison with the seismic one. We focus our study to the zone located above 0.96 $\Rs$, that is the zone where changes in the subsurface stratification have been found by \citet{Lefebvre05} along the solar cycle. The upper limit is fixed at 0.998 $\Rs$, beyond which the superadiabatic zone extends, the turbulence acts strongly and the rotation changes quickly. We have chosen  the zone where the $f$-modes have a good sensitivity. 
At a first glance, we  notice that most of these variables present non negligible variations in the studied region. 

We first comment on the common trends of both figures:
\begin{itemize}
	\item The variations shown on the different panels are dominated by the change in radius imposed in the calibration and at the second order by the change of composition for set 3.
%	The change in luminosity result either from the radius change via the Stefan's law (set 1), or from the calibration by the evolution code (set 2 and 3), i.e. from variations in the nuclear core contrary to the cycle change which probably comes from more external layers. 
	\item The way we introduce the change in luminosity acts on the nuclear burning layers and has a negligible effect on the subsurface layers. So for clarity, we have not plotted the models 2 and 3 of set 1, nor 5 and 8 of set 2, in fact their curves are aligned with the models having the same radius in Figure \ref{F-param}. In contrast the observed luminosity variation of $10^{-3}$ along the solar cycle is most likely coming from the very external layers. In fact, the variation of radius and of the photospheric temperature of our models produces a change of luminosity determined by the Stefan's law. It is generally too small to exhibit any structural effect, except for models 10 and 13 where the variation of luminosity coming from the variation of radius and temperature is near from the imposed variation of luminosity.
	\item The behaviors of $c$, $H_p$ and $H_{\rho}$ are very similar i.e. a bump around 0.99 $R_{\odot}$ with opposite variations between models of different radius; this bump exists also for the temperature. %which slightly decreases toward the surface. 
This position corresponds to the transition between neutral He and He$^+$ as discussed in \citet{Lopes97}.
	\item The differences of $\Gamma_1$ and $\nabla_{real}$ have similar variations and present a double peak near 0.99 $\Rs$ with almost equal amplitude. In addition to the bump at 0.99 $\Rs$, the second bump could be due to the transition H$^+$/H. It is reasonable that in this convective region, the gradient of the structure follows the adiabatic exponent.
	\item The variations of $\kappa$ and $\nabla_{rad}$ present a sign change at 0.99 $\Rs$, connected to the variation of the opacities in the region where the light elements are partially ionized. The variation of the pressure and density modifies the corresponding opacity coefficients.
\end{itemize}	

Figure \ref{F-param2} differs nevertheless from Figure \ref{F-param}. Figure \ref{F-param} shows two symmetrical groups of models depending on the value of the calibrated radius. In the set 3 another change  comes from the modification of the heavy elements contribution (about 10\%) which induces a small change in helium (about 3\%) and hydrogen. In order to separate properly the effect of radius from the effect of composition, we have calculated model ($M_{14} $) that includes only the change in composition. We show in Figure \ref{F-param2} the changes induced by the composition effect. As an evident consequence, we loose the symmetry between the group of models with a greater radius compared to the models with a smaller one, this is mainly visible on the temperature and $\Gamma_1$ differences.

These first behaviors indicate that the subsurface layers above 0.96 $\Rs$ are significantly affected by a change in radius and composition. To go further, we shall estimate the differences in $f$-mode frequencies issued from these models.

\subsection{Calculation of  the theoretical $f$-modes for the different models}
			\label{S-fmodes}

\begin{figure*}[htbp]
	\centerline{\includegraphics[angle=-90,width=8cm]{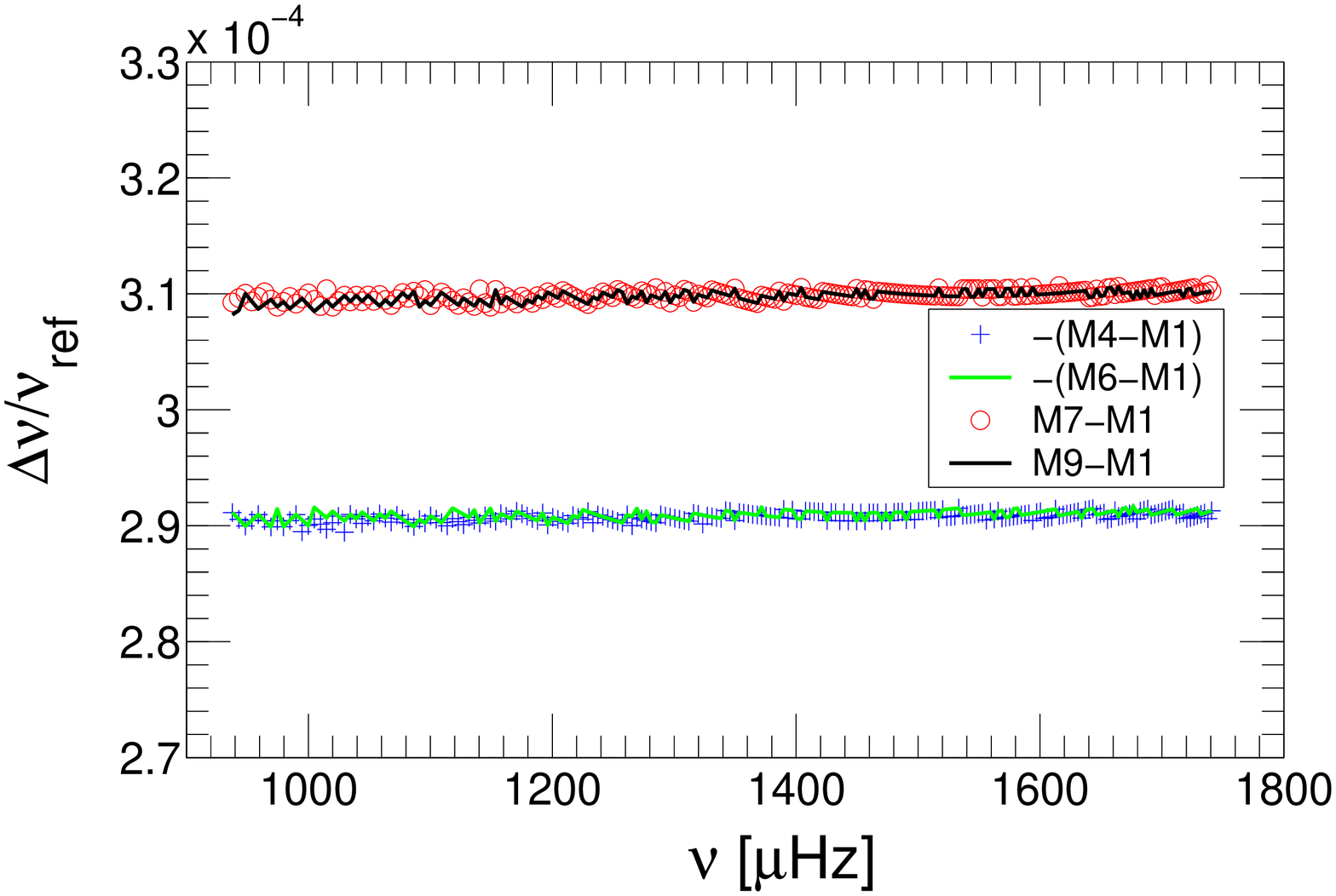}
							\includegraphics[angle=-90,width=8cm]{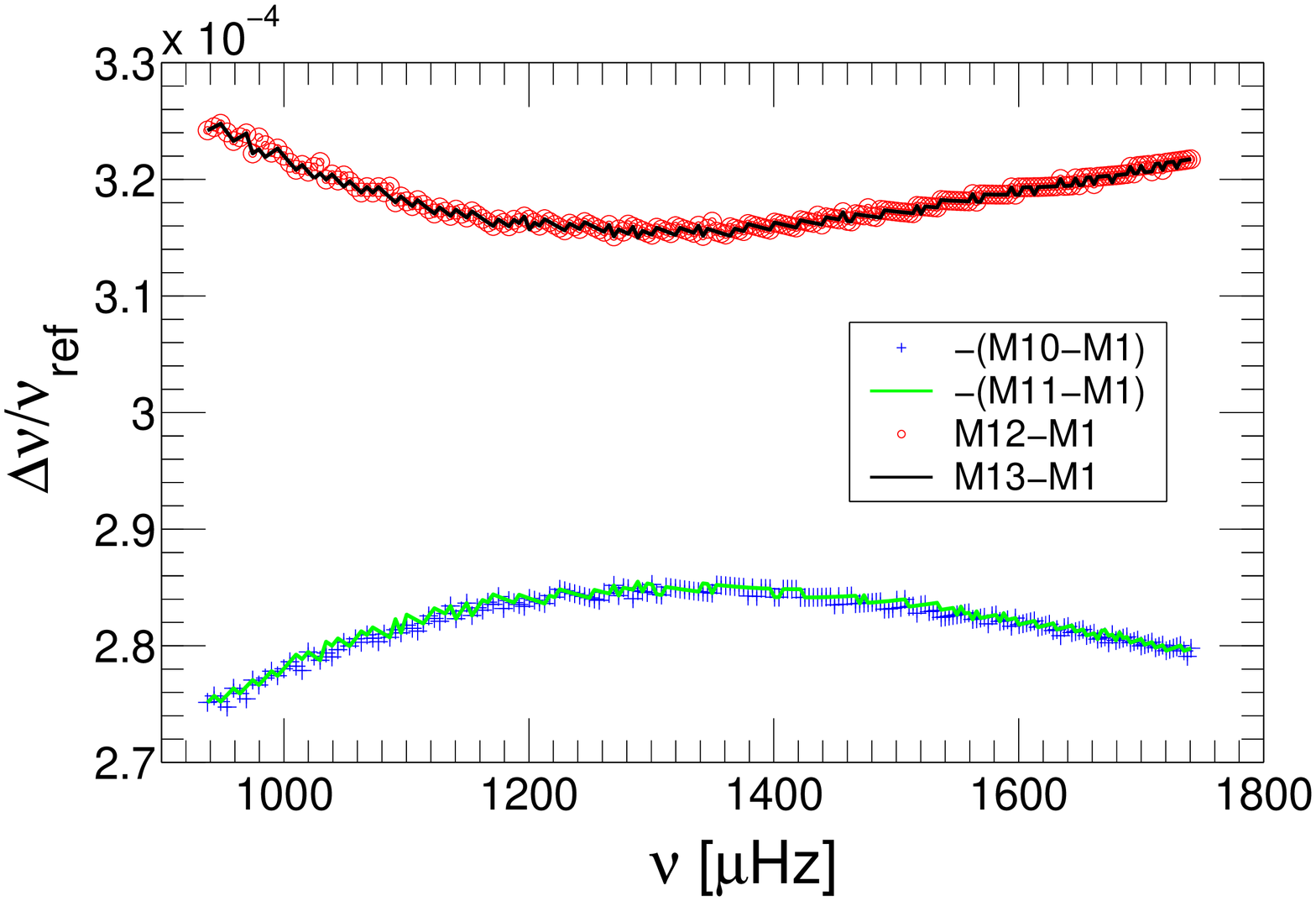}}
	\caption{Relative difference of $f$-modes frequencies between models of set 2 (left panel) or models of set 3 (right panel) and $M_1$ versus the absolute frequencies. We do not plot the difference between the two reference models $M_{14}$ and $M_1$ because this difference  is of the order of $10^{-5}$.}
	\label{F-fmodes}
\end{figure*}

We compute  for each model the theoretical $f$-mode frequencies using the  oscillation code ADIPLS \citep{Christensen82} to see the impact of the different changes on these quantities. Figure \ref{F-fmodes} shows the relative difference of frequencies between each model and the reference model $M_1$ versus the corresponding absolute frequencies.

We note first that the $f$-mode frequencies of a model with a larger radius are smaller than those of the reference model.  The change of $2\times10^{-4}$ in radius leads to a relative change of about $3\times10^{-4}$ on $f$-mode frequencies.

The left panel of Figure 4 corresponds to sets 1 and 2. The difference associated to $M_4$ and $M_6$ is slightly bigger in absolute value from the difference associated to $M_7$ and $M_9$ due to the value of the reached real radius: +135 km for $M_4$ and $M_6$ and -143 km for $M_7$ and $M_9$ (this difference is within the precision imposed of the model radius). The frequency dependence has almost a flat behavior, whereas this is not the case for set 3 in the right panel where the relative difference has an extremum around $\approx 1300 \, \mu Hz$ due to the additional effect of the change in composition, which produces non-uniform changes as already discussed.
\clearpage
% Figure 4
\clearpage

\section{Numerical inversion of theoretical $f$-mode frequency to infer subsurface stratification changes}
			\label{S-inversion}
			
In this section we do the inversion of the $f$-mode frequencies for degrees $l$ between 100 and 300, in a range slightly bigger than the one chosen previously with the corresponding solar observational quantities \citep{Lefebvre05} to see the reproductivity of the radial variation for known models, to qualify the procedure and to extrapolate a new estimate of the radius variation along the solar cycle.

%\subsection{Reconstruction of the subsurface layers stratification}
			%\label{S-results}

We use the same formalism as in \citet{Lefebvre05} to infer the changes in the position of subsurface layers from the $f$-mode frequency variations. A relation between the relative frequency variations $\delta\nu/\nu$ for $f$-modes and the associated Lagrangian perturbation of the radius $\delta r/r$ of the subsurface layers has been established by \citet{Dziem04}:
\begin{equation}
\left(\frac{\delta\nu}{\nu}\right)_l=-\frac{3l}{2\omega^2 I}\int
dI\frac{g}{r}\frac{\delta r}{r} \label{Eq-eq_radius}
\end{equation}
where $l$ is the degree of the $f$-modes, $I$ is the moment of inertia as classically defined in \citet{Dziem04}, $\omega$ is the angular frequency of the eigenpulsation ($\omega=2\pi\nu$) and $g$ is the gravity acceleration.
The validity of this equation is limited to the case where any magnetic field effect is explicitly introduced in the equations \citep{Dziem04} which is the case in this work. It also supposes that, if $y$ and $z$ are respectively the vertical and horizontal eigenfunctions, the property of $f$-modes $y^2 = l(l+1)z^2$ is satisfied.

%The validation of this equation has been defined by \inlinecite{Dziem04}: \textbf{it does not include any magnetic field effect which is our case; moreover if $y$ and $z$ are respectively the vertical and horizontal eigenfunctions,  $y^2=l(l+1)z^2$ for $f$-modes, and if we note ${\cal E} = y^2+l(l+1)z^2$, then both assumptions, i.e. Equation 43 of their article and $2\Lambda yz \approx l{\cal E}$ where $\Lambda=l(l+1)$, are completely valid when using $f$-modes.}  %\footnote{By checking the calculations of \inlinecite{Dziem04}, we notice a typo error in the equation expressing ${\cal D}_r$. There is a problem of homogeneity and the right one is ${\cal D}_r=2\Gamma\Lambda z\lambda+6\frac{gr\rho}{p}(y^2-\Lambda yz)$ where \Lambda=l(l+1).} 
%\end{enumerate}

This equation allows us to obtain $\delta r$ from $\delta\nu/\nu$. For these inversions, we used $M_1$ as the reference model %calibrated to the seismic radius $\Rs=6.9599\times10^5$ km, 
and a standard Regularized Least-Square technique, since Equation \ref{Eq-eq_radius} defines an ill-posed inverse problem \citep{Tikhonov77}). The validity of this equation for the different cases analyzed here will be discussed in the last section. 

\clearpage
% Figure 5
\begin{figure*}[htbp]
	\centerline{\includegraphics[angle=-90,width=8cm]{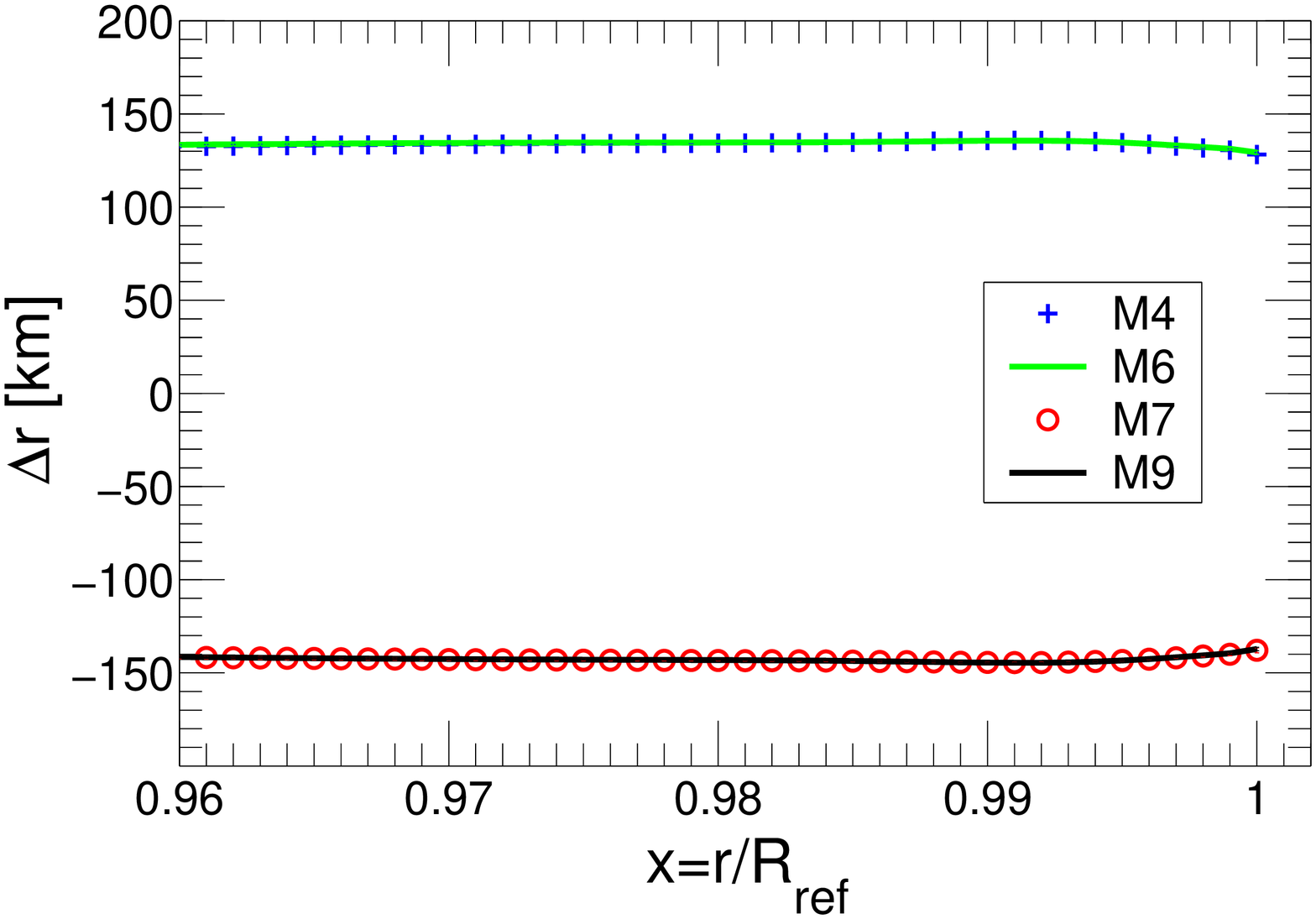}
							\includegraphics[angle=-90,width=8cm]{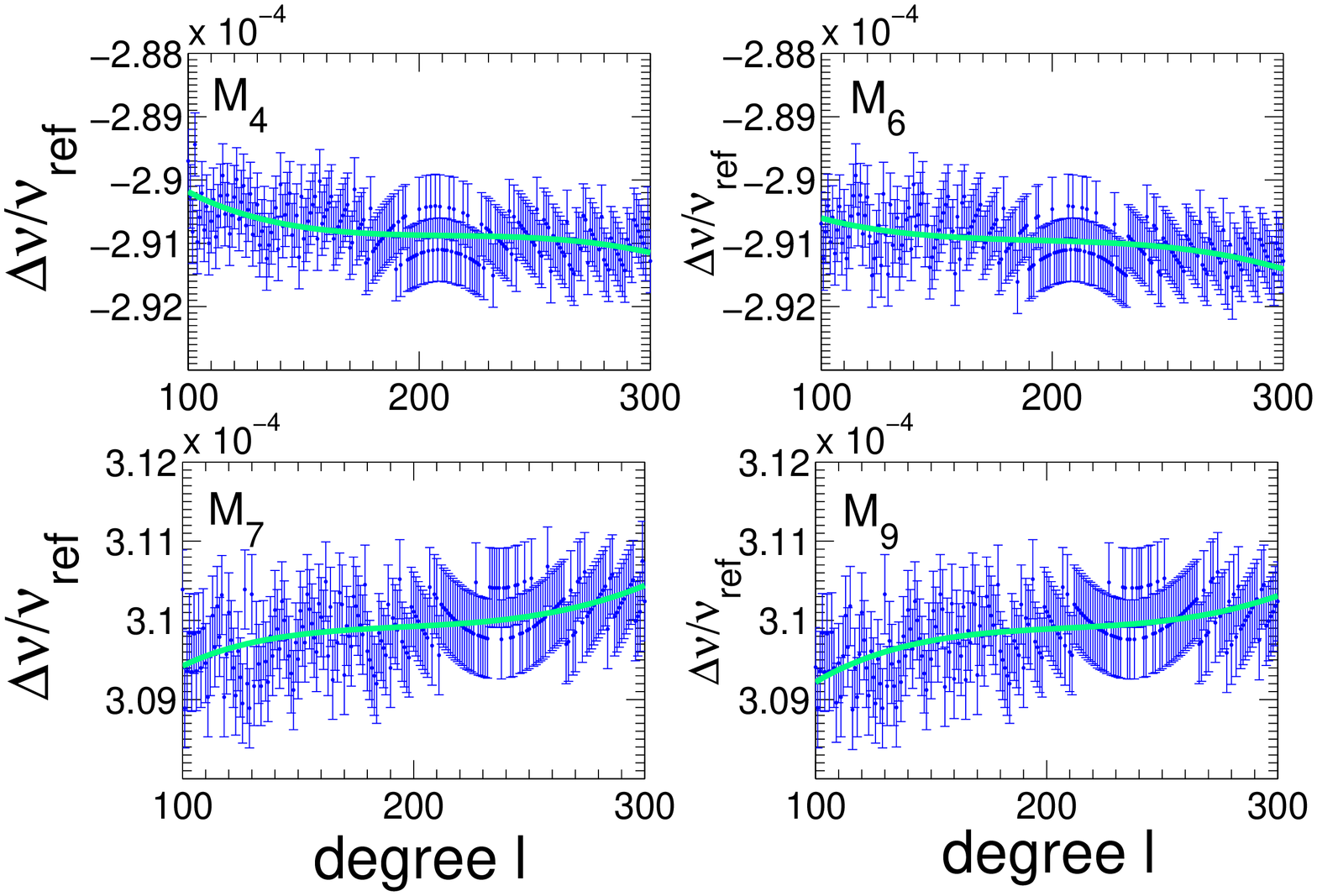}
						 }
	\caption{\textit{Left panel}: Radial variation $\Delta r$ as a function of the fractional radius $x=r/R_{ref}$, obtained by inversion of Equation \ref{Eq-eq_radius} with frequency variations represented in the left panel of Figure \ref{F-fmodes} for set 2. The error bars are the standard deviation after averaging over a set of random noise added to the relative frequencies reconstructed in the right panel. 
\textit{Right panel}: For each model, relative variation of frequencies $\Delta\nu/\nu_{ref}$ as a function of the degree $l$. The points with error bars (that have been arbitrarily added to the data, such as $\sigma=0.5\times10^{-6}$) represent the input data, and the solid curves are the result of direct integration of Eq. \ref{Eq-eq_radius}, providing the solutions in the left panel.}
	\label{F-inversion1}
\end{figure*}

% Figure 6
\begin{figure*}[htbp]
	\centerline{\includegraphics[angle=-90,width=8cm]{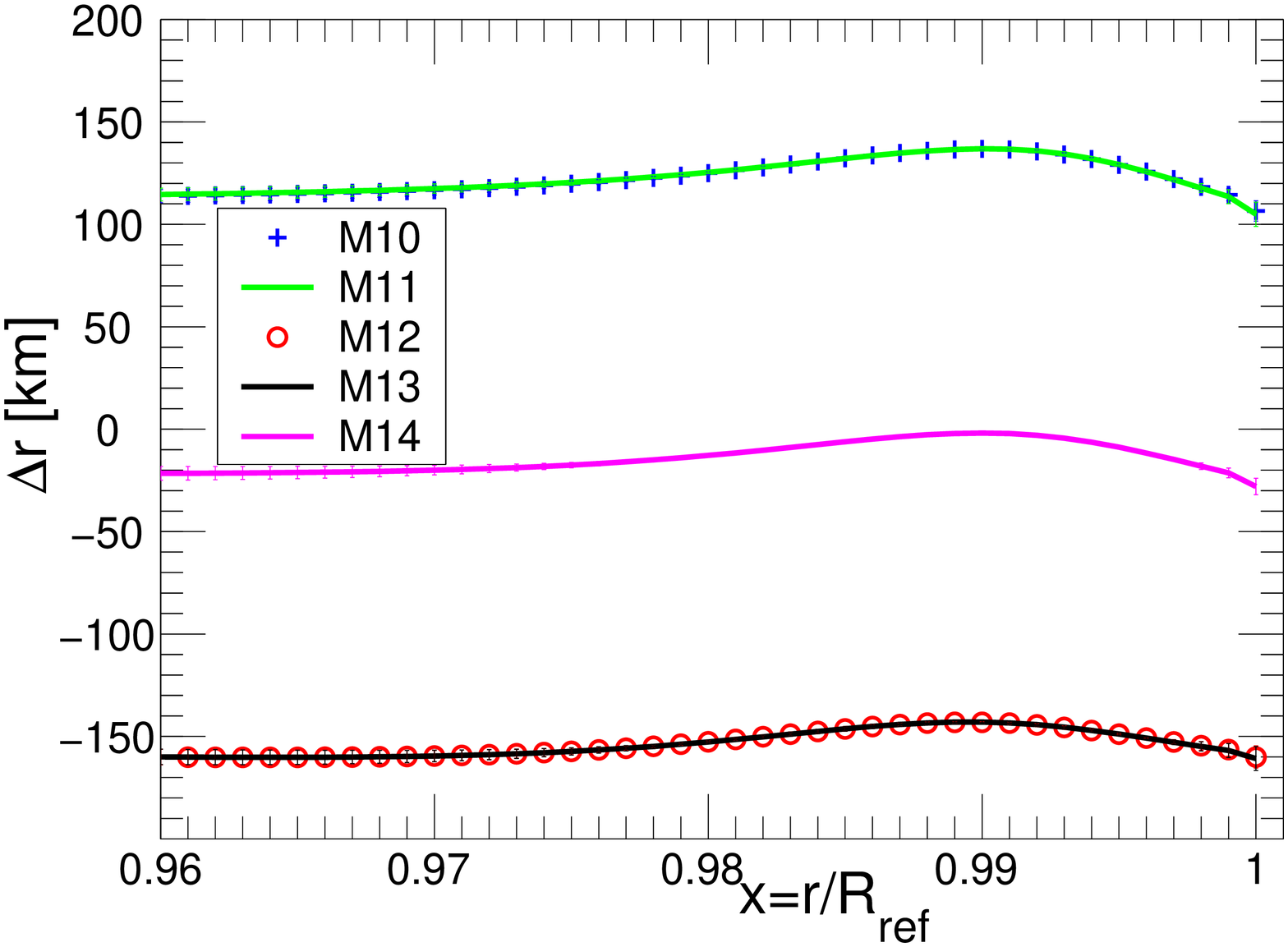}
							\includegraphics[angle=-90,width=8cm]{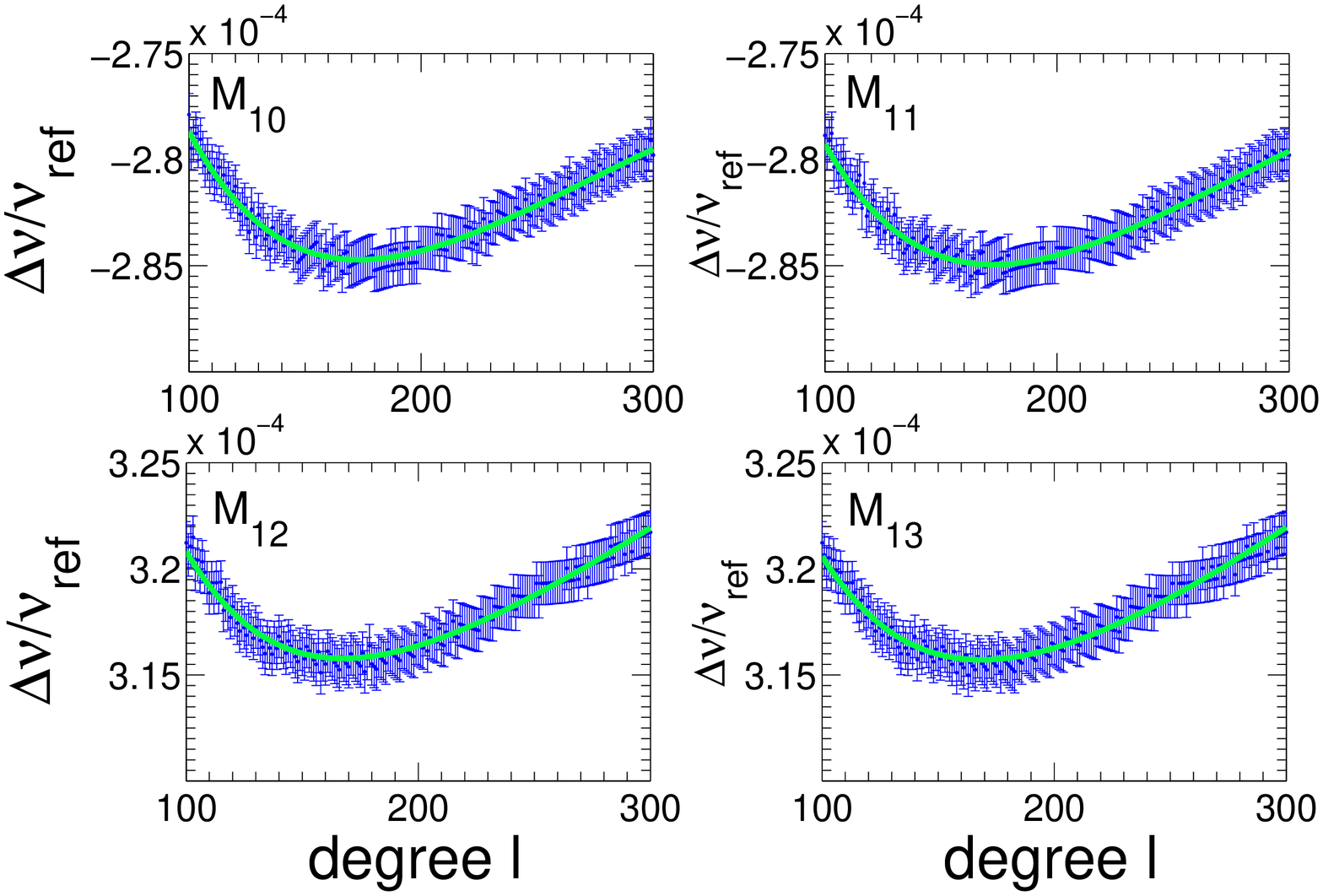}
						 }
	\caption{Idem than Figure \ref{F-inversion1} for models of set 3. Error bars have been arbitrarily added to the data, such as $\sigma=1\times10^{-6}$. Note here the behavior of the curves near $x=0.99$ reflecting a non-uniform change in the subsurface stratification. For clarity we do not plot the relative variation of frequencies for model $M_{14}$ that is 10 times smaller than for the other models.}
	\label{F-inversion2}
\end{figure*}
\clearpage

Figures \ref{F-inversion1} and \ref{F-inversion2} show results from inverting Equation \ref{Eq-eq_radius} for each computed model. These figures also show the frequencies computed by integrating the solutions using Equation \ref{Eq-eq_radius}, and how well they match  the model frequencies within the error bars. The uncertainties were set arbitrarily to $0.5\times10^{-6}$ and $1\times10^{-6}$ respectively for the need of the inversion process and the will to mimic real variations. This fact  guarantees the good quality of the inversion.
The main characteristics of our solutions are:
\begin{itemize}
		\item The inversion solutions plotted in Figure \ref{F-inversion1} for set 2 lead to an almost uniform radial variation.  The amplitude of the radial variation is similar to the one imposed at the surface, as expected from the relative variations of frequencies in the left panel of Figure \ref{F-fmodes}, i.e. about $\pm140$ km with a sign respecting the nominal one. Nevertheless, there is a small difference near the surface (see also next section), which is due to the poor spatial resolution of the kernels at the surface. This problem has already been raised in \citet{Lefebvre05} where the shape of the kernels is shown.
    \item The solutions for set 3 shown in Figure \ref{F-inversion2} are slightly different. Below $0.98\Rs$, the variation is constant but closer to the surface there are non-monotonic changes in the stratification with a bump centered at $0.99\Rs$. The uncertainty of the localization of this bump is governed by the characteristic width of our kernels that is about $0.005\Rs$. 
    \item These variations are different in shape and in amplitude to those found by \citet{Lefebvre05}. This is not surprising because we do not yet introduce the dynamical processes which generate the solar cycle. %But the problem here is different as we used models without differential rotation nor magnetic field, contrary to the reality where all these processes are mixed. We recall that this is a first step in the understanding of the dynamical processes governing the stratification of the subsurface layers.
    \item Since the only difference between some models of set 2 and some others of set 3 is the different value of $\left(Z/X\right)_{sf}$,  this bump is clearly a consequence of the introduction of this change in composition affecting the subsurface layers through the pressure and mainly the equation of state. We note that the value of  $\Delta$R is not any longer about 140 km but about 110 km (135-20 km) for models $M_{10}$ and $M_{11}$ and -160 km (142+20 km) for $M_{12}$ and $M_{13}$ at $0.96\Rs$, this is easily explained by looking at the $\Delta$R between the two reference models of about 20 km at this depth. It is clear that a change of composition has an effect largely below the superficial layers.
\end{itemize}

In fact the inversion process implies that the solution is not unique. By adjusting the regularisation parameters, we can find another profile with two bumps for $\Delta r$ with also a very good fit to frequency. That solution has not been retained because (i) the variations at the surface are not in agreement with the input radius variation as cited in Table \ref{T-table_models}, (ii) the error bars are bigger and (iii) the curves are more oscillating. Moreover, we cannot presently give a physical interpretation to this solution. On the contrary, for the first solution, we discuss in section \ref{S-discussions} how we understand the present inversions.

\section{Discussion}
	\label{S-discussions}
	
%In this section, we investigate the physical meaning of the radial displacement $\Delta r$ shown in Figures \ref{F-inversion1} and \ref{F-inversion2}. 
Equation 8 of \citet{Dziem04} supposes that the radial position $r$ used in Equation \ref{Eq-eq_radius} is the lagrangian radius. $f$-modes are surface oscillation trapped waves, so each mode oscillates around an equilibrium radius. We relate in this section this radial position to that of the structure.

\subsection{ Validity of the procedure of inversion}

The extraction of the solar subsurface structure and the photospheric radius variation along the solar cycle is not so easy to determine and several papers have been published with different conclusions \citep{Dziem01, Antia03, Sofia05}. The most recent work of \citet{Dziem04} shows the difficulty to obtain a general expression for the inversion, so this work allows to test such a procedure. We verify in this section if the displacement $\Delta r$ of Equation \ref{Eq-eq_radius} corresponds to the radial displacement at a given mass for the different cases studied.

\clearpage
% figure 7
\begin{figure*}[htbp]
	\centerline{\includegraphics[width=8cm]{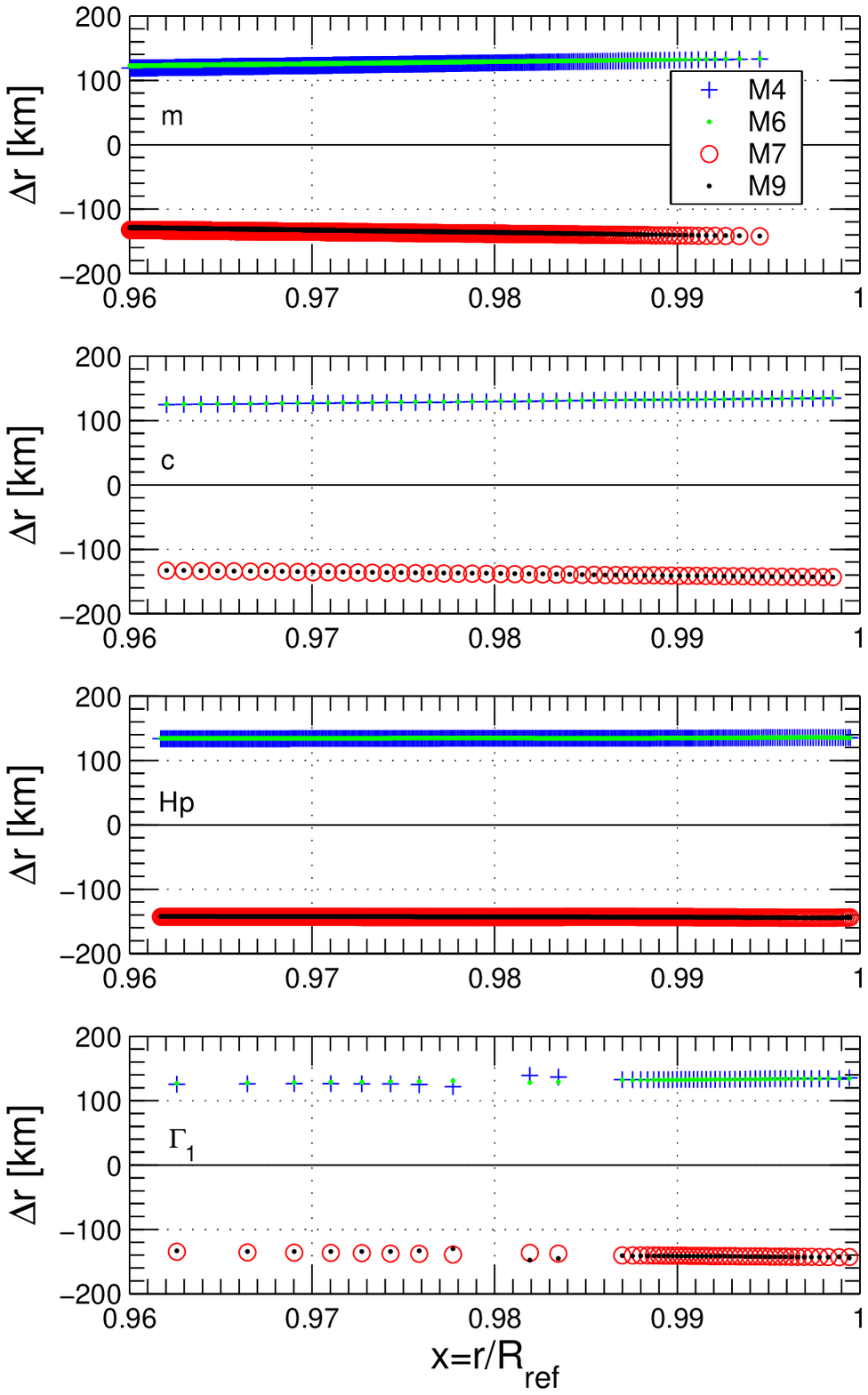}
							\includegraphics[width=8cm]{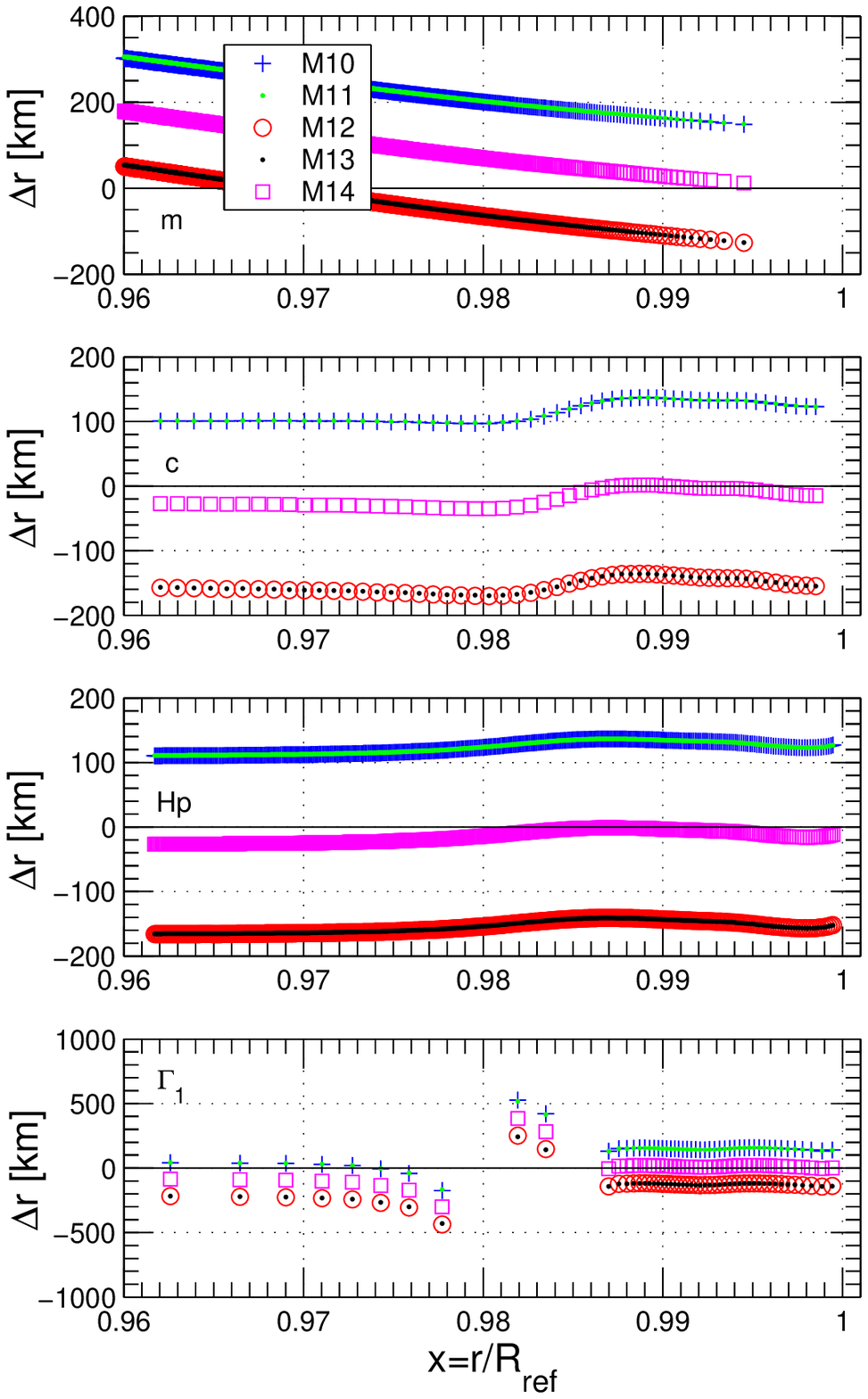}}
	\caption{\textit{Left panel}: Radial displacement $\Delta r$ of a given m(r), p(r), $c$,$H_p$, $\Gamma_1$ and $\kappa$ defined in the reference model ($M_1$) for models of set 2 (the curves of  set 1 models superpose the previous ones). These curves have to be compared to the left panel of Figure \ref{F-inversion1}. \textit{Right panel}: Idem with models of set 3.}
%These curves have to be compared with the ones obtained in Figure \ref{F-inversion2}.
	\label{F-radius_param}
\end{figure*}

% figure 8
\begin{figure}[htbp]
	\centerline{\includegraphics[angle=-90,width=8cm]{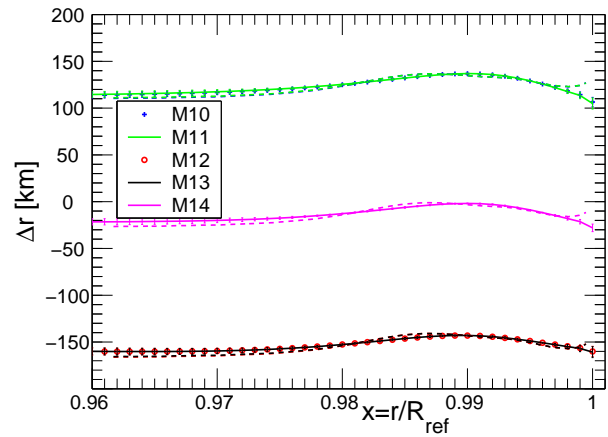}}
	\caption{Radial displacement $\Delta r$ issued from the inversion (same curves than in the left panel of Figure \ref{F-inversion2} added with the associated displacement at $H_p$ constant as plotted in the right panel of Figure \ref{F-radius_param} (here dashed curves associated at the same color for the corresponding model). Note the good superposition of the different curves.}
%These curves have to be compared with the ones obtained in Figure \ref{F-inversion2}.
	\label{F-superposition}
\end{figure}
\clearpage

With this purpose, we calculate for each set (sets 1 and 2 lead to the same result) the radial displacement of a layer corresponding to a given value of the structural quantities, i.e. $\Delta r = (r_N-r_1)_{q_N=q_1}$ where $q$ represents a model structure quantity like $m$, $c$, $H_p$ or $\Gamma_1$, and $N$ is the model number.

Figure \ref{F-radius_param} shows the panels representing $\Delta r$ at constant $m$, $c$, $H_p$ and $\Gamma_1$. For sets 1 or 2, we have the same behaviour for all the quantities and Equation \ref{Eq-eq_radius} can be applied without any doubt. For the set 3, the radial difference $\Delta r$, computed by the procedure described above, differs from one quantity to another. The only curve similar in amplitude and in shape with the one computed by inversion of $f$-mode frequencies in Figure  \ref{F-inversion2} is the curve computed at fixed pressure height $H_p$. Effectively Figure \ref{F-superposition} shows an excellent superposition of the corresponding curve and the radial displacement $\Delta r$ issued from the inversion. This agreement, in shape and amplitude, demonstrates that the validity of equation 1 in this case requires to associate the radial displacement to this quantity. It is in fact not surprising because the $f$-mode frequencies are naturally sensitive to the pressure scale height.

We note also that the inversion is not able to reproduce precisely the behavior very near the surface due to the lack of sensitivity of  the $f$-modes. But we can deduce from this study an estimate of the uncertainty on the determination of the photospheric radius variation: we estimate it of the order of 15\%.

\subsection{A new prediction on the solar radius variation along the solar cycle}

We show in this paper that a pure variation in the solar radius produces changes in the subsurface layers that are in the same zone than that studied with the observed $f$-modes by \citet{Lefebvre05}. Such a change is characterized by changes in the subsurface stratification and more exactly variations in the computed position of the layers. We show that the changes detected by $f$-modes are physically related to the variation of the pressure. In our modelling we are able to produce non-uniform changes by slightly modifying the chemical composition below the solar surface. This  is not  excluded if one introduces dynamical processes in keeping the same constraints on luminosity and radius. The two studies show different stratifications of the outlayers but with similar effects on the $f$-mode frequencies. Considering larger effects on the $f$-mode frequencies than in the solar case, we propose to deduce from  this study an estimate of the variation of the solar radius along the cycle. From $\Delta\nu/\nu$ of about $3\times10^{-4}$ for a $\Delta r$ of about 140 km, we lead to a solar radius variation along the solar cycle of about $7\pm 1$ km for the observed variations in frequency of about $1.5\times10^{-5}$. This result is consistent with the radius extrapolated by \citet{Lefebvre05} (see respectively Figures 1 and 3 of this article). It is interesting to note that this radius variation estimate is also compatible (in order of magnitude) with the observed low degree acoustic mode variation of about 0.4 $\mu Hz$ along the solar cycle of the low degree $p$-modes \citep{Fossat87,Chaplin01}. 

\subsection{Perspectives}

This study represents the first step on the way to explain the variations of the subsurface stratification along the solar cycle and reinforces the interest of the $f$-modes. The next step will be the development of models including magnetic field and other dynamical processes, which will permit a more realistic study.  Nevertheless it will suppose also an improved expression for the inversion of $f$-modes which is not yet obtained. We believe that the introduction of a magnetic field will influence the stratification of the subsurface layers. The study by \citet{Nghiem06} pointed out non-uniform radial changes, depending on the importance of the magnetic pressure. Moreover, it will be also interesting to take into account the differential rotation and consequently some asphericity to confront them to the subsurface latitudinal stratification over the 11-year cycle. 

We would like to emphasize that a better knowledge of these subsurface layers will contribute to our understanding (i) of the dynamics of the 11-year solar cycle and (ii) of the Sun-Earth relationship for space climate. This supposes in parallel  the simultaneous measurement of the radius and frequencies change. This study is in the framework of coming future space missions: SDO (see \url{http://sdo.gsfc.nasa.gov/}) and PICARD (see \url{http://smsc.cnes.fr/PICARD/} and \citet{Thuillier06}). 

The DynaMICCS/HIRISE perspective \citep{Turck06,Turck08}, proposed in the framework of the ESA Cosmic Vision 2015-2025, could bring even more data on all sources of internal variability by putting together instruments which will follow seismically all the layers down to the core together with instruments measuring the variability of the above atmosphere. 

\acknowledgments
We would like to thank two anonymous referees who have helped us to clarify our results and to enrich this paper. S. Lefebvre was supported by a CNES/GOLF postdoctoral grant in SAp/CEA.

\end{document}